\documentclass[fleqn,usenatbib]{mnras}

\usepackage{newtxtext,newtxmath}

\usepackage[T1]{fontenc}

\DeclareRobustCommand{\VAN}[3]{#2}
\let\VANthebibliography\thebibliography
\def\thebibliography{\DeclareRobustCommand{\VAN}[3]{##3}\VANthebibliography}

\usepackage{graphicx}	
\usepackage{amsmath}	
\usepackage{subcaption}
\usepackage{bm}

\title[Radius inflation of WASP-76b]{Evidence of Radius Inflation in Radiative GCM Models of WASP-76b due to the Advection of Potential Temperature}

\author[F. Sainsbury-Martinez et al.]{Felix Sainsbury-Martinez,$^{1}$\thanks{E-mail: f.sainsbury-martinez@leeds.ac.uk}
Pascal Tremblin,$^{2}$
Aaron David Schneider,$^{3,4}$
Ludmila Carone,$^{5,6}$\newauthor
Isabelle Baraffe,$^{7,8}$
Gilles Chabrier,$^{8,7}$
Christiane Helling,$^{9,5}$
Leen Decin$^{4}$ and
Uffe Gr\r{a}e J\o{}rgensen$^{3}$
\\
$^{1}$School of Physics and Astronomy, University of Leeds, Leeds LS2 9JT, UK\\
$^{2}$Universite Paris-Saclay, UVSQ, CNRS, CEA, Maison de la Simulation, 91191, Gif-sur-Yvette, France\\
$^{3}$Centre for ExoLife Sciences, Niels Bohr Institute, {\O}ster Voldgade 5, 1350 Copenhagen, Denmark\\
$^{4}$Institute of Astronomy, KU Leuven, Celestijnenlaan 200D, 3001 Leuven, Belgium\\
$^{5}$Space Research Institute, Austrian Academy of Sciences, Schmiedlstrasse 6, 8042 Graz, Austria\\
$^{6}$Centre for Exoplanet Science, School of Physics \& Astronomy, University of St Andrews, North Haugh, St Andrews KY169SS, UK \\
$^{7}$University of Exeter, Physics and Astronomy, EX4 4QL Exeter, UK\\
$^{8}$\'Ecole Normale Sup\'erieure, Lyon, CRAL (UMR CNRS 5574), Universit\'e de Lyon, France\\
$^{9}$Institute for Theoretical Physics and Computational Physics, Graz University of Technology, Petersgasse16/II, 8010 Graz, Austria\\
}

\date{Accepted 2023 June 21. Received 2023 June 16; in original form 2023 April 27}

\pubyear{2023}

\begin{document}
\label{firstpage}
\pagerange{\pageref{firstpage}--\pageref{lastpage}}
\maketitle

\begin{abstract}
Understanding the discrepancy between the radii of observed hot Jupiters and standard 'radiative-convective' models remains a hotly debated topic in the exoplanet community. One mechanism which has been proposed to bridge this gap, and which has recently come under scrutiny, is the vertical advection of potential temperature from the irradiated outer atmosphere deep into the interior, heating the deep, unirradiated, atmosphere, warming the internal adiabat, and resulting in radius inflation. Specifically, a recent study which explored the atmosphere of WASP-76b using a 3D, non-grey, GCM suggested that their models lacked radius inflation, and hence any vertical enthalpy advection. Here we perform additional analysis of these, and related, models, focusing on an explicit analysis of vertical enthalpy transport and the resulting heating of the deep atmosphere compared with 1D models. Our results indicate that, after any evolution linked with initialisation, all the WASP-76b models considered here exhibit significant vertical enthalpy transport, heating the deep atmosphere significantly when compared with standard 1D models. Furthermore, comparison of a long time-scale (and hence near steady-state) model with a Jupiter-like internal-structure model suggests not only strong radius-inflation, but also that the model radius, $1.98 \mathrm{R_{J}}$, may be comparable with observations ($1.83\pm0.06 \mathrm{R_{J}}$). We thus conclude that the vertical advection of potential temperature alone is enough to explain the radius inflation of WASP-76b, and potentially other irradiated gas giants, albeit with the proviso that the exact strength of the vertical advection remains sensitive to model parameters, such as the inclusion of deep atmospheric drag. 
\end{abstract}




\section{Introduction}

Observations of hot Jupiters \citep{2011ApJ...729L...7L} and hot brown dwarfs (see Fig. 4 of \citealt{2020MNRAS.499.5318C}) have revealed a significant discrepancy between standard `radiative-convective' single column (1D) atmospheric models and the properties of observed objects: observed radii of highly irradiated objects tend to be significantly larger than 1D atmospheric models suggest (see, for example, Figure 1 of \citealt{Komacek_2017}). This indicates that said 1D models are likely failing to capture some key physics or dynamics which drive the observed radius discrepancy (i.e. inflation). In 1D models this discrepancy is `solved' via the inclusion of an intrinsic/internal temperature, which essentially acts to heat the deep atmosphere (internal adiabat) to a more physical value, allowing for atmospheric retrievals of transit observations, without actually elucidating on exactly what dynamics drives this heating other than typically claiming that it is linked with thermal escape from the interior \citep{2002A&A...385..156G,2003A&A...402..701B,2003ApJ...588.1121S,2004ApJ...603L..53C,2019ApJ...884L...6T}.\\
In a collective effort to understand this deep heating/radius inflation problem, a vast array of different physical mechanisms have been suggested as possible causes/solutions (see \citealt{Baraffe_2009,2010SSRv..152..423F,2014prpl.conf..763B} for a more in-depth overview of many of the proposed mechanisms) including tidal heating and dissipation \citep{2010ApJ...714....1A,2019MNRAS.484.5845L}, the ohmic dissipation of electrical/magnetic energy \citep{Batygin_2010,2010ApJ...719.1421P,2012ApJ...750...96R,2021A&A...648A..80H,2022A&A...658L...7K}, the deep deposition of kinetic energy \citep{2002A&A...385..156G}, enhanced opacities which inhibit interior cooling \citep{Burrows_2007}, double-diffusive convection which hampers convective heat transport \citep{2007ApJ...661L..81C}, or the vertical advection of potential temperature (first proposed and studied in 2D by \citealt{2017ApJ...841...30T} and studied in 3D by \citealt{2019A&A...632A.114S,2021A&A...656A.128S}). \\
Fortunately, observations can help us to narrow down which of the above mechanisms might be responsible for the observed radius inflation. Specifically, observational studies of hot Jupiters and hot brown dwarfs (e.g. \citealt{Demory_2011,2011ApJ...729L...7L,2016ApJ...818....4L,2018A&A...616A..76S,2020MNRAS.499.5318C}) have revealed a clear trend: a general increase in the observed radius of highly irradiated gaseous planets with stellar irradiation, except in the case of very-highly irradiated objects in very short orbits (e.g. SDSS1411B - \citealt{2018MNRAS.481.5216C}) where little to no inflation is observed. One such mechanism which can explain this trend without the inclusion of model-dependent fine tuning is the vertical advection of potential temperature (i.e. enthalpy). \\
Briefly this mechanism can be understood as follows: for a tidally locked, gaseous exoplanet, the strong stellar irradiation leads to a very hot outer atmosphere paired with a very strong super-rotating equatorial jet. This driving can be understood via a 2D stationary circulation model, in which, due to mass and angular momentum conservation, significant vertical winds arise (as proposed/seen in \citealt{2017ApJ...841...30T,2019A&A...632A.114S}). Note that such a view is opposed by \citet{2011ApJ...738...71S}, who assume/propose that only the irradiated layers of the atmosphere are meteorologically active, and that deeper layers are either quiescent or purely convective {(not that the latter would have any negative implications for our mechanism, beyond the temperature of the adiabat)}. Not only do our results disagree with this view (see \autoref{fig:nominal_zonal_means}), but other studies, such as \citet{2020MNRAS.496.3582C,2022A&A...664A..56S} have shown that significant wave activity and zonal/vertical winds can occur in these deep atmospheric layers. If this holds true, and we propose that it does, these aforementioned vertical winds carry high potential temperature fluid parcels from the hot (radiative) outer atmosphere deep into the interior (where radiative effects tend to zero - as shown in \autoref{fig:enthalpy}), driving the formation of a non-convective (i.e. advective) adiabat at lower pressures than 1D models (without an artificially increased internal/intrinsic temperature) would predict. {Because this adiabat forms at lower pressures , and because the radiative, advective and deep convective (i.e. interior) regions must smoothly connect, the internal adiabats temperature temperature is significantly increased when compared to a model which lacks advection and considers a radiative-convective boundary alone. In turn, this increase in the temperature of the internal adiabat, leads to an increase in the internal entropy, and hence an inflated radius. \\}
This is very similar to what occurs in a 1D model when the internal/intrinsic temperature is increased{, although here it is occurring due to fundamental physics.}
An example of this can be seen in \citet{2019ApJ...884L...6T}, who find a clear link between the pressure of the RCB (radiative-convective boundary), i.e. where the outer atmosphere connects with the interior adiabat, and the intrinsic temperature, i.e. internal heat flux that their models impose. 
However the heating which drives the formation of this non-convective adiabat has nothing to do with heat transport from the interior. Rather it is heating associated with the irradiated outer atmosphere, which should, at steady-state, balance any outwards heat transport from the interior, stalling any internal cooling and leading to a net zero internal flux (i.e. no heating from the interior), a stable, inflated, radii, and a natural link between radius inflation and surface irradiation. \\

It is important to note that this mechanism is distinct from the kinetic energy transport and deposition mechanism proposed by \citet{2002A&A...385..156G}. In their mechanism, stellar irradiation is converted to kinetic energy in the outer atmosphere (by atmospheric pressure gradients), this energy is then somehow transported down towards the interior (possibly by, for example, Kelvin-Helmholtz instabilities, vertical advection, or waves), where it then dissipates, heating the deep atmosphere and warming the internal adiabat. Rather, the mechanism we (and \citealt{2017ApJ...841...30T,2019A&A...632A.114S}) propose skips these uncertain energy conversion steps, and instead we directly transport hot (high enthalpy) material from from the outer atmosphere to the deep atmosphere via already present flows and circulations. \\

Recently, \citet{2022A&A...666L..11S}, called into question the validity of vertical potential temperature advection as a possible explanation for the radius inflation of the ultra-hot Jupiter WASP-76b (\citealt{2016A&A...585A.126W,2019A&A...623A.166S,2020Natur.580..597E,2022AJ....163..107K}), arguing that their (hot-start) 3D atmospheric models, calculated with expeRT/MITgcm, and including a self-consistent, non-grey radiative transfer model (see \citealt{2022A&A...664A..56S} for a detailed discussion of this code), suggested that coupling between radiation and dynamics alone is not sufficient to explain the inflated radii of highly-irradiated, gaseous, exoplanets. \\

Here, we intend to investigate this claim in more detail, performing additional analysis of the nominal WASP-76b simulation  discussed in \citet{2022A&A...666L..11S} along with additional, cooler-start (i.e. cooler initial deep adiabats) calculations that were run exclusively for this work. Specifically, we intend to investigate the vertical mass and enthalpy (i.e. potential temperature) transport in these models, confirming if vertical advection plays a significant role in the dynamics, before comparing the steady-state 3D models with internal-structure models based upon the work of \citet{2010RPPh...73a6901B} in order to confirm how much, if any, of the inflated radius of WASP-76b potential temperature advection alone can explain. \\
In \autoref{sec:Methods}, we start with a brief overview of expeRT/MITgcm before introducing the models discussed as part of this work. This is followed, in \autoref{sec:Results} with our analysis, focusing on the vertical transport of potential temperature and its implications for the steady-state deep atmosphere of our WASP-76b models. We finish, in \autoref{sec:Discussion} by discussing the implications of our results, with a particular focus on the sustainability of potential temperature advection as an explanation for the inflated radii of highly irradiated, tidally locked, gaseous exoplanets. 

\section{Methods} \label{sec:Methods}
The methodology and models used in this work are based on the work of \citet{2022A&A...664A..56S} and \citet{2022A&A...666L..11S}. Here we give a brief overview of the GCM used to calculate the WASP-76b models considered here, before giving a more in depth description of said WASP-76b models setup. 

\subsection{expeRT/MITgcm}

Briefly, expeRT/MITgcm \citep{2020MNRAS.496.3582C,2022A&A...664A..56S} builds on the dynamical core of the MITgcm \citep{2004MWRv..132.2845A}, pairing said core with the petitRADTRANS \citep{2019A&A...627A..67M} radiative transfer (RT) model in order to enable the long model integration times required to explore the steady-state dynamics of the deep atmospheres of hot Jupiters, whilst also maintaining the accuracy of a multi-wavelength radiation scheme. \\
expeRT/MITgcm solves the primitive equations of meteorology \citep{Vallis17,Showman_2009}, for an ideal gas, on an Arakawa C-type cubed sphere (designed to avoid numerical issues near the poles which occur due to singularities in the coordinate system; for more details of this grid, see, for example, \citealt{https://doi.org/10.1002/qj.49711046321}) with a horizontal resolution C32\footnote{{C32 is comparable to a resolution of 128 × 64 in longitude and
latitude}} and a vertical grid that contains a combination of 41 linearly in $\log(P)$ (i.e. log-pressure) spaced layers between $1\times10^{-5}$ and 100 bar, paired with 6 linearly in $P$ spaced layers between 100 and 700 bar. {As in \citet{Showman_2009,2020MNRAS.496.3582C} this model includes a horizontal fourth-order Shapiro filter (with $\tau$ = 25 s) in order to smooth grid-scale noise.}
Additionally, expeRT/MITgcm includes a linear Rayleigh-drag {(which is also known as a linear-basel drag scheme - see \citealt{2020MNRAS.496.3582C}, particularly Section 2.3 and Appendix A for a discussion of this dynamics preserving approach as well as comparisons with other drag-schemes)} at the bottom of the atmosphere ({between $490$ and $700$ bar}) and a sponge layer at the top of the atmosphere (for $P<1\times10^{-4}$ bar). We discuss the implications of this Rayleigh-drag on the vertical advection of potential temperature, and hence radius inflation, in more detail in \autoref{sec:Results}. {Note: we selected 700 bar as the maximum pressure of our simulation domain in-order to balance modelling a sufficient portion of the deep atmosphere with the increasing computational costs of modelling high-pressure regions (due to their increased dynamical timescales).}  \\
Radiatively, the outer atmosphere is heated and cooled using a runtime (i.e. coupled), multi-wavelength, RT scheme based upon petitRADTRANS. Specifically, the radiative dynamics are updated every 100 seconds, quadruple the dynamical time step ($\Delta t_{\textrm{dy}}=25$ s), with the radiative transport calculated using a correlated-k approach that includes 5 wavelength bins each of which contain 16 Gaussian quadrature points (see \citealt{1989JQSRT..42..539G} for an introduction to the correlated-k approach to RT, {and Appendix B of \citealt{2022A&A...664A..56S} for a discussion of the accuracy of the limited wavelength bin approach}). Note that opacities for the RT scheme are based on a pre-calculated pressure-temperature grid, assume local chemical equilibrium, and include the following gas absorbers (with data taken from the ExoMol\footnote{\url{www.exomol.com} and \citet{2016JMoSp.327...73T,2021A&A...646A..21C}} database): $\mathrm{H}_2\mathrm{O}$, $\mathrm{CO}_2$, $\mathrm{CH}_4$, $\mathrm{NH}_3$, $\mathrm{CO}$, $\mathrm{PH}_3$, $\mathrm{H}_2\mathrm{S}$, $\mathrm{TiO}$, $\mathrm{VO}$, $\mathrm{HCN}$, $\mathrm{Na}$, $\mathrm{K}$ and {$\mathrm{FeH}$}. Additionally, the RT model includes Rayleigh-scattering for both $\mathrm{H}_2$ and He, and collision-induced-absorption for $\mathrm{H}_2-\mathrm{H}_2$, $\mathrm{He}-\mathrm{He}$ and {H$^{-}$} (see \citealt{2022A&A...666L..11S} for more details). {We do not include equilibrium condensation since, assuming that the latent heat release is low \citep{2003A&A...399..297W,2019JPhCS1322a2028H,2019A&A...626A.133H}, it should have little effect on the photosphere, especially for WASP-76b whose day-side can be assumed to be cloud free, and is simply too hot for condensation to occur.} \\

Finally, the inclusion of an artificial Rayleigh-drag scheme in the deep atmosphere implies that an additional energy source term must be added to the deep atmosphere to account for the conversion of energy lost from drag to heat (\citealt{2013ApJ...764..103R,2020MNRAS.496.3582C,2022A&A...664A..56S}), {which is then locally returned to the atmosphere}. This takes the form:
\begin{equation}
 	\frac{dT}{dt} = \frac{u^{2}}{c_{p}\tau_{\mathrm{drag}}},
\end{equation}
where $T$ is the local temperature of the atmosphere, $u$ is the horizontal {(zonal plus meridional)} wind speed, $\tau_{drag}$ is the Rayleigh-drag timescale at the bottom of the atmosphere, and $c_{p}$ is the heat capacity at constant pressure. 

\subsection{Models of WASP-76b}
WASP-76b is a tidally locked ultra-hot Jupiter-like planet ($M=0.92\pm0.03\mathrm{M_{J}}$) that orbits its host star at a distance of 0.033 AU, corresponding to an orbital period of 1.81 Earth days, and which appears to exhibit significant radius inflation, with an observed radius of $1.83\pm0.06\mathrm{R_{J}}$ \citep{2016A&A...585A.126W}. The host star, WASP-76, is a hot yellow-white (F7V) main-sequence star with an effect temperature of $T_\mathrm{eff}=6250\pm100$ K and a radius of $R_{*}=1.73\pm0.04\mathrm{R_{\odot}}$ \citep{2018A&A...616A...1G}. Further, all our models assume a fixed specific heat capacity, $c_{p}=13784$ $\mathrm{J\,kg^{-1}K^{-1}}$and a fixed specific gas constant, $R=3707$ $\mathrm{J\,kg^{-1}K^{-1}}$, which corresponds to a adiabatic index $\gamma\simeq1.36$ (these values have been extracted from petitRADTRANS). However the Rayleigh-drag timescale does vary, with the majority of our models setting $\tau_\mathrm{drag}=1$ day, and a low-drag model setting $\tau_\mathrm{drag}=1000$ days. Finally, we include zero heat flux from the interior, meaning that any deep atmospheric heating is purely due to downwards enthalpy advection from the irradiated outer atmosphere. \\

\begin{figure} %
\begin{centering}
\includegraphics[width=1.0\columnwidth]{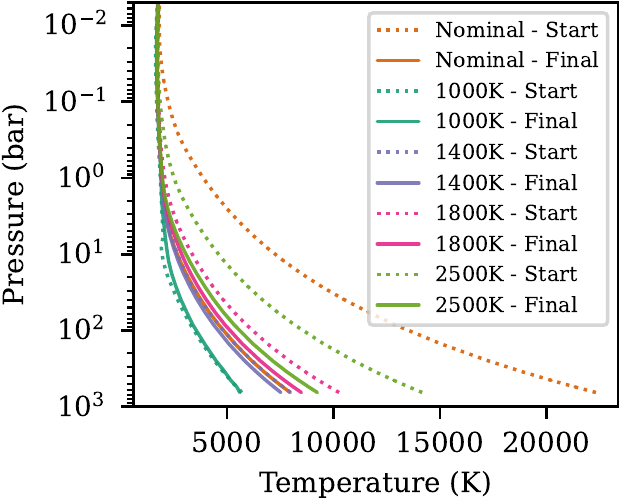}
\caption[Start and Final T-P profiles for all models with different deep initialisation temperatures]{ Horizontal mean Temperature-Pressure profiles for our five WASP-76b atmospheric models with different initialisation temperatures. For each of the variable initial temperature models considered here, i.e. the nominal (4000 K), 1000 K, 1400 K, 1800 K, and 2500 K start models, we include a profile near initialisation (dotted) and a profile at the end of the models runtime (solid). Note that the nominal model has been run for significantly longer than the other models (\autoref{tab:table1}), and hence is likely to represent the steady state that all aforementioned models are converging towards.  \label{fig:T_evo_comp} }
\end{centering}
\end{figure}

Here we consider 6 models of WASP-76b, five of which only differ in the temperature profile used to initialise them, and one in which the strength of the deep Rayleigh-drag has been reduced (as previously mentioned). For the former models, the initialisation profile is a combination of an isotherm, based upon the stellar irradiation, in the outer atmosphere (i.e. for $P<1$ bar), and an adiabat, with a reference temperature ($\theta$) taken at 1 bar, throughout the deep atmosphere (i.e. $P>10$ bar), with a linear interpolation between the two profiles between 1 and 10 bar. Here we consider reference temperatures of $\theta=4000$ K (i.e. the nominal model which was first presented in \citealt{2022A&A...666L..11S}, but which has been further evolved as part of this work), 2500 $\mathrm{K}$, 1800 $\mathrm{K}$, 1400 $\mathrm{K}$ and 1000 $\mathrm{K}$, which range from hotter than the adiabat of the final nominal model of \citet{2022A&A...666L..11S} to cooler - thus allowing us to explore models in which the deep atmosphere is both heating and cooling. These initial profiles can be seen in \autoref{fig:T_evo_comp}, where we plot the initial profile of each variable initialisation model as a dashed line. On the other hand, the low-drag model (with $\tau_\mathrm{drag}=1000$ days) is initialised from a snapshot of the nominal model (with $\theta=4000$ K) taken after 40,000 days of simulation time. Note that, other than the nominal model, all the models featured here were performed as part of this work. 

\begin{table}
\begin{tabular}{l|l|l|l|l}
Model & Run Time (d) & Peak $F_\mathcal{H}$ (erg cm$^{-2}$) & Mean $F_\mathcal{H}$ (erg cm$^{-2}$)  \\ \hline
Nominal & 155,100 & $-8.40\times10^{11}$ & $-6.24\times10^{7} $ \\
2500 K & 57,100 & $-1.81\times10^{12}$  &$ -2.35\times10^{7}$   \\
1800 K & 69,200 & $-8.70\times10^{11}$ & $-2.99\times10^{7} $ \\
1400 K & 69,700 & $-5.09\times10^{11}$ & $-6.04\times10^{7} $ \\
1000 K & 69,100 & $-2.14\times10^{11}$ & $-8.59\times10^{7} $ \\
Low Drag & 50,000 &$-2.46\times10^{12}$ & $-3.84\times10^{8} $
\end{tabular}
\caption{Peak (downward) and global mean vertical enthalpy flux for five WASP-76b models in which either the temperature of the initial deep adiabat or the strength of the deep drag have been changed, along with the nominal model presented in \citet{2022A&A...666L..11S}. Note: the Low Drag model has been run for 10,000 additional days using a snapshot of the nominal model after 40,000 days (i.e. the 'evolved' model of \citealt{2022A&A...666L..11S}) of simulation time as an initial condition. Further, the mean vertical enthalpy flux for the nominal model at an equivalent timestep to that of the Low Drag model remains essentially unchanged. }
\label{tab:table1}
\end{table}
\begin{figure*} %
\begin{centering}
\begin{subfigure}{0.99\columnwidth}
\begin{centering}
\includegraphics[width=1.0\columnwidth]{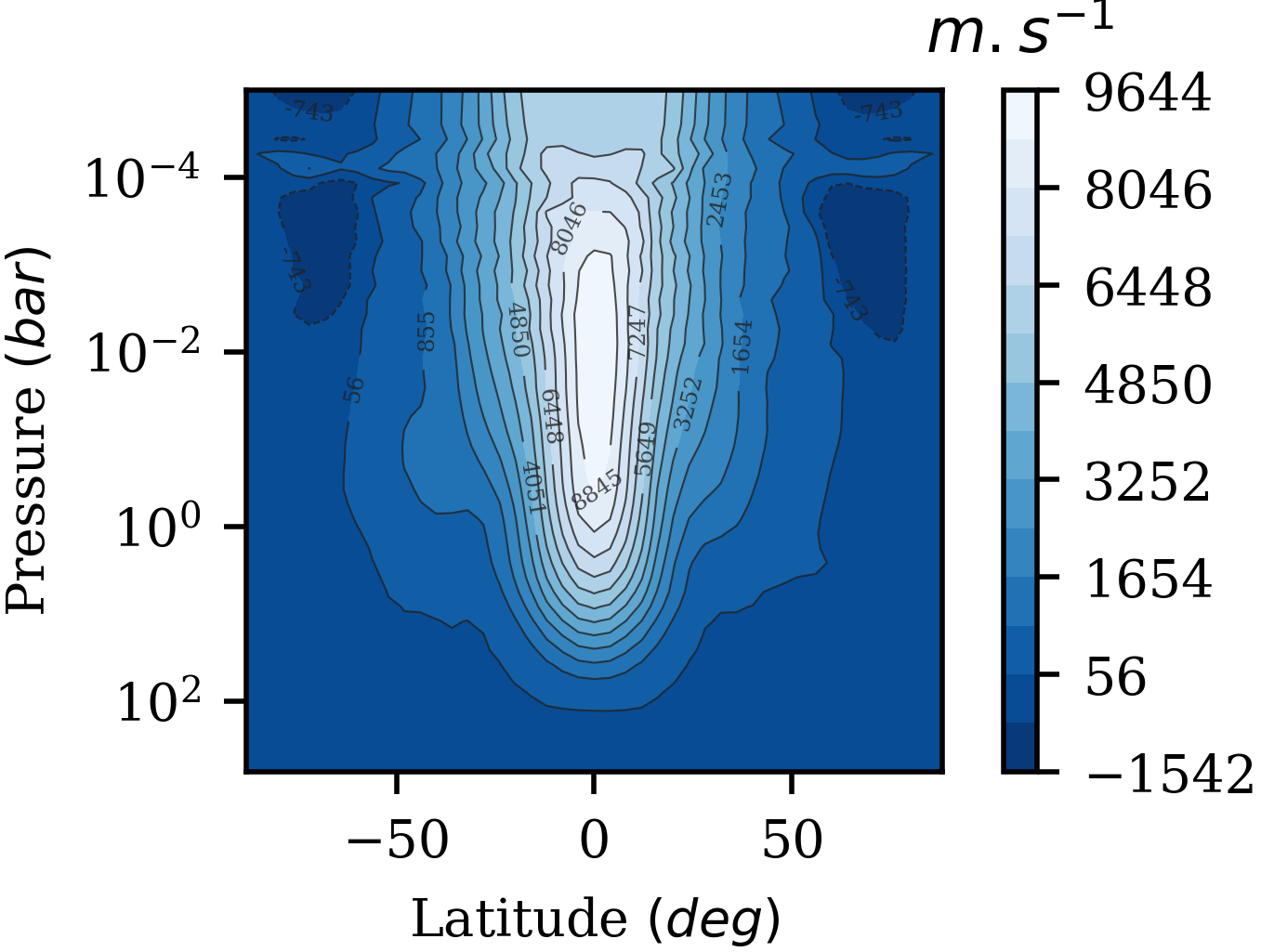}
\caption[]{Zonal-Mean Zonal Wind  \label{fig:zonal_wind} }
\end{centering}
\end{subfigure}
\begin{subfigure}{0.99\columnwidth}
\begin{centering}
\includegraphics[width=1.0\columnwidth]{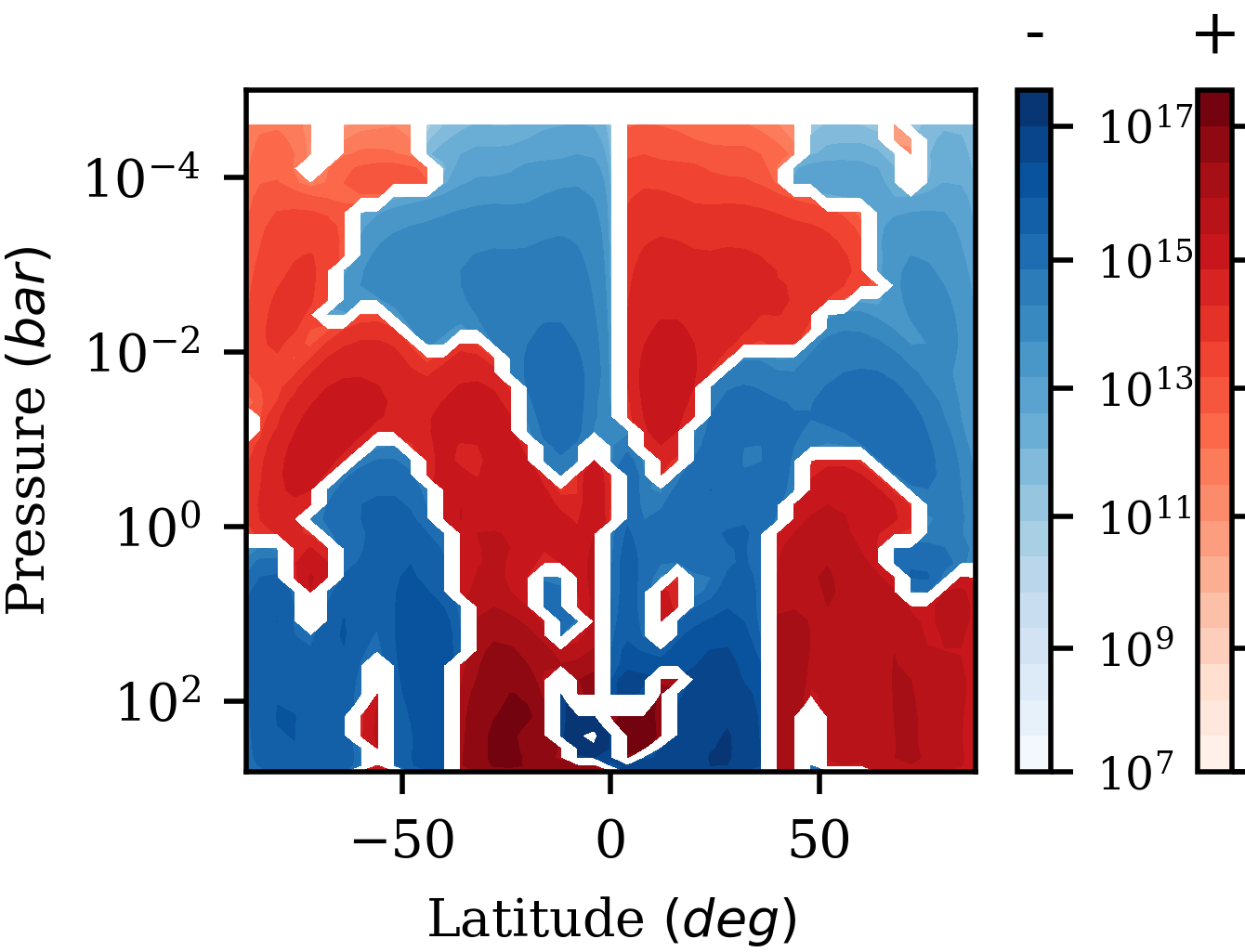}
\caption[]{Meridional Mass Streamfunction  \label{fig:meridional_streamfunction} }
\end{centering}
\end{subfigure}
\caption[Zonal Wind and Meridional Streamfunction for Nominal Model]{The zonal-mean zonal wind (left) and meridional circulation streamfunction (right) for the nominal GCM model of WASP-76b. In the zonal wind profile, easterly winds are positive and westerly winds are negative, whilst in the meridional circulation profile, we plot the streamfunction using a log scale in order to clearly illustrate the full circulation profile, especially in the outer atmosphere. Here, clockwise circulations are shown in red and anti-clockwise in blue - these circulations combine to reveal an equatorial upwelling in the outer atmosphere driven by the strong day-side irradiation, and an equatorial downflow in the deep atmosphere, which is linked with the downward advection of potential temperature. \label{fig:nominal_zonal_means} }
\end{centering}
\end{figure*}
\begin{figure*} %
\begin{centering}
\begin{subfigure}{0.315\textwidth}
\begin{centering}
\includegraphics[width=0.99\columnwidth]{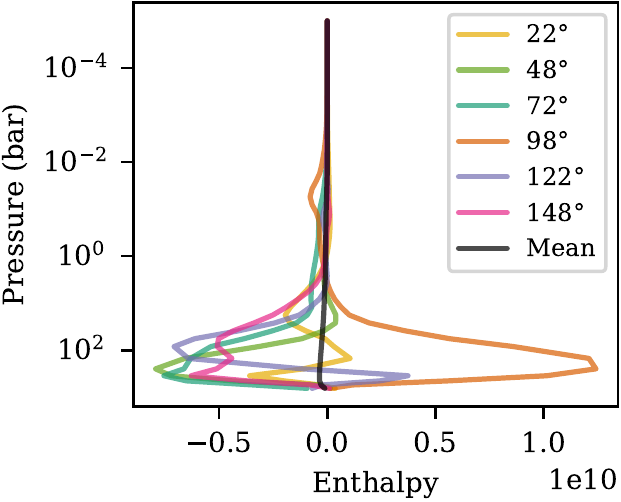}
\caption[]{1000 K - Longitudinal Variations  \label{fig:Enthalpy_Profiles_1000K_slices} }
\end{centering}
\end{subfigure}
\begin{subfigure}{0.315\textwidth}
\begin{centering}
\includegraphics[width=0.99\columnwidth]{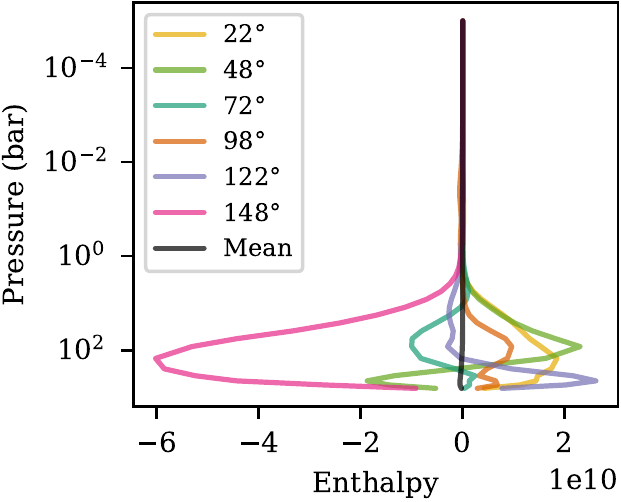}
\caption[]{Nominal - Longitudinal Variations   \label{fig:Enthalpy_Profiles_nominal_slices} }
\end{centering}
\end{subfigure}
\begin{subfigure}{0.315\textwidth}
\begin{centering}
\includegraphics[width=0.99\columnwidth]{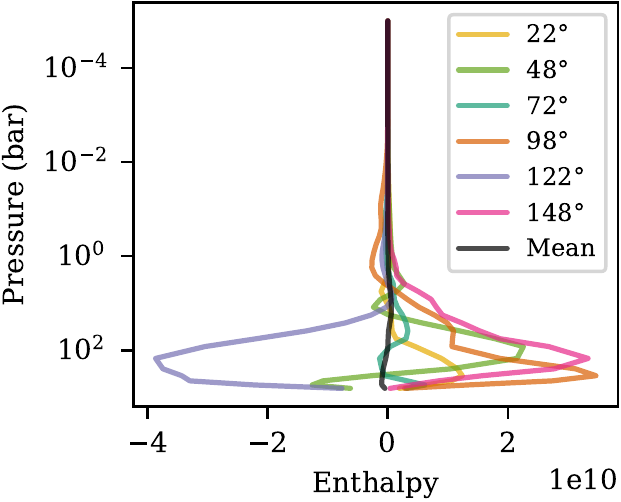}
\caption[]{2500 K - Longitudinal Variations   \label{fig:Enthalpy_Profiles_2500K_slices} }
\end{centering}
\end{subfigure}
\begin{subfigure}{0.315\textwidth}
\begin{centering}
\includegraphics[width=0.99\columnwidth]{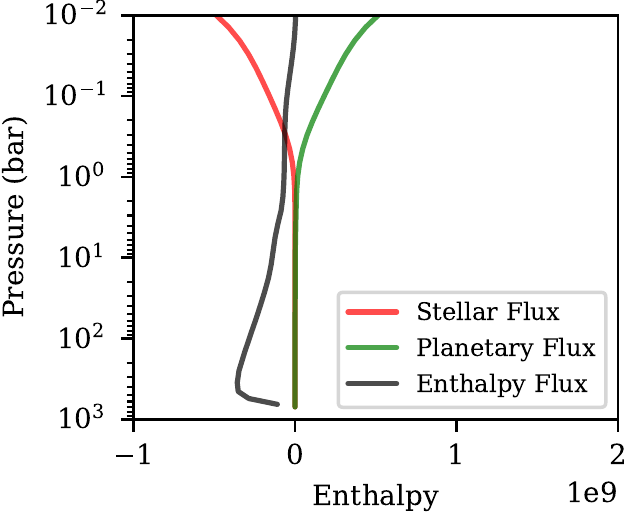}
\caption[]{1000 K - Global Mean  \label{fig:Enthalpy_Profiles_1000K_mean} }
\end{centering}
\end{subfigure}
\begin{subfigure}{0.315\textwidth}
\begin{centering}
\includegraphics[width=0.99\columnwidth]{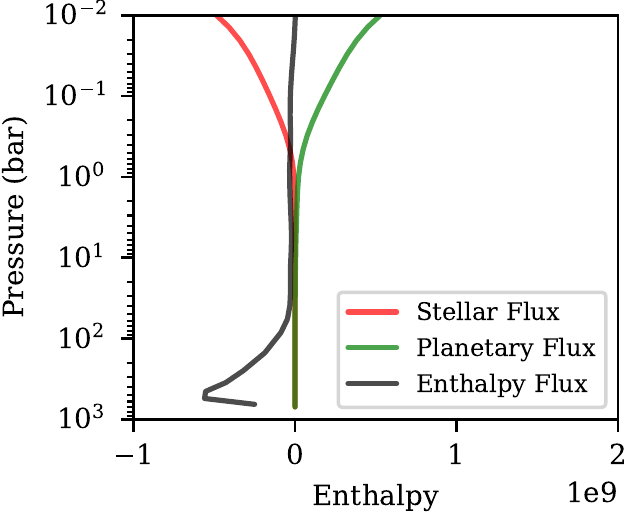}
\caption[]{Nominal - Global Mean   \label{fig:Enthalpy_Profiles_nominal_mean} }
\end{centering}
\end{subfigure}
\begin{subfigure}{0.315\textwidth}
\begin{centering}
\includegraphics[width=0.99\columnwidth]{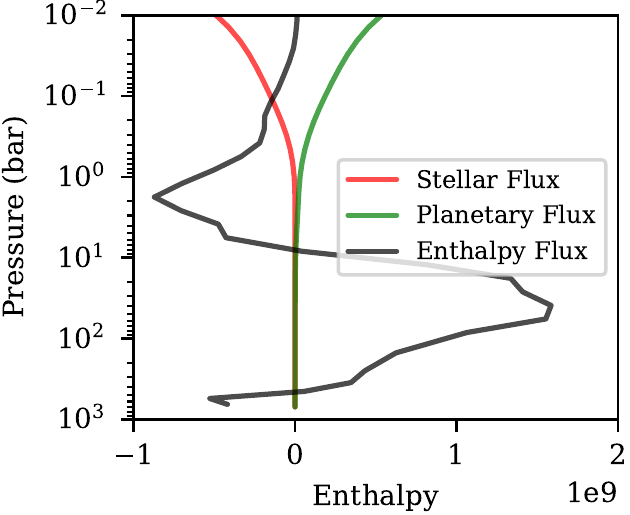}
\caption[]{2500 K - Global Mean   \label{fig:Enthalpy_Profiles_2500K_mean} }
\end{centering}
\end{subfigure}
\caption[Enthalpy transport profiles for three WASP-76b models]{Longitudinal variation of the meridional mean (top) and global mean (bottom) vertical enthalpy flux $F_\mathcal{H}$ profiles for three WASP-76b GCM models with different initial deep adiabat temperatures; 1000 K - left, nominal (4000 K) - middle, and 2500 K - right. In the top row figures, we plot the vertical enthalpy flux profiles at 6 different longitudes, ranging from just east of the anti-stellar point to just west of the substellar point, as well as the global mean vertical enthalpy flux profile. However, since the mean flux is significantly smaller than the local fluxes, we replot the mean profiles in the bottom row in order to better demonstrate the vertical variations in enthalpy transport, focusing on the advection into the deep atmosphere. Here we also include the horizontal mean stellar (incoming) and planetary (outgoing) fluxes in order to reinforce that the deep atmosphere is radiatively quiescent. Note: the 2500 K profiles are calculated near the start of the simulation when the cooling is strongest - similar results can be found near initialisation for the nominal and other hot-start models. \label{fig:enthalpy} }
\end{centering}
\end{figure*} 

\begin{figure} %
\begin{centering}
\includegraphics[width=1.0\columnwidth]{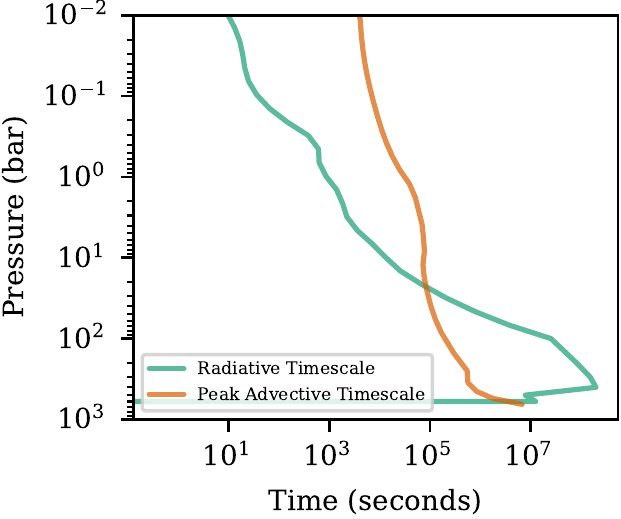}
\caption[]{ Comparison of the peak vertical advective time-scale ($\tau_{adv}=H/u_{r}$, where $H$ is the atmospheric scale height and $u_{r}$ is the maximum downward velocity) and the global-mean radiative timescale for the near steady-state nominal model. Note how, despite both timescales increasing with pressure, the rapid increase in optical depth means that advection dominates over radiation in the deep atmosphere.  \label{fig:Time_Comp} }
\end{centering}
\end{figure}

\begin{figure} %
\begin{centering}
\includegraphics[width=1.0\columnwidth]{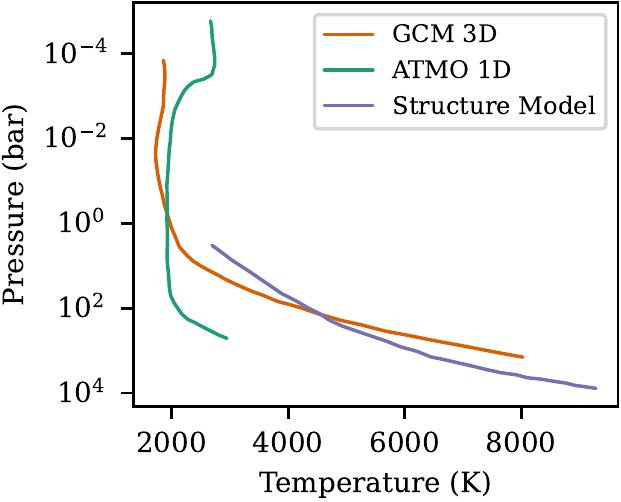}
\caption[Nominal Model T-P Profiles including comparison with ATMO and Baraffe Structure Models]{ Comparison of the horizontal mean Temperature-Pressure profile from the nominal GCM model of WASP-76b (orange) with both a 1D model calculated using ATMO (with $T_\mathrm{int}=100$ K - green) and an internal-structure model, based on the work of \citet{2003A&A...402..701B,2004ApJ...603L..53C}, of a hot Jupiter with a mass of $0.9\mathrm{M_{J}}$ and an inflated radius of $1.98\mathrm{R_{J}}$ (purple). \label{fig:T_comp_1D_structure} }
\end{centering}
\end{figure}

\section{Results} \label{sec:Results}

A broader analysis of the nominal model, after 86,000 days of runtime, is presented in \citet{2022A&A...666L..11S}. Instead, here, we focus our analysis on the vertical advection of potential temperature, including what drives this advection, what effect it has on the deep atmosphere, and how much, if any, of WASP-76b radius inflation can be attributed to it. \\

We start our analysis with the nominal model, which, after over 155,000 days of simulation time {(which corresponds to over 10,000 advective turnover timescales in the deep atmosphere - see \autoref{fig:Time_Comp})}, is approaching steady-state at almost all simulated pressures. Here we find that the strong day/night temperature difference associated with the combination of both tidal-locking and a hot host-star has resulted in the formation of a rapid super-rotating jet (see \autoref{fig:zonal_wind}, which plots the zonal mean zonal wind at 155,000 days) that extends significantly into the deep atmosphere: at the equator the region in which $u_\mathrm{zonal}>1000$ $\mathrm{m\,s^{-1}}$ extends to pressures greater than 10 bar. Such deep jets were already predicted in \citet{2020MNRAS.496.3582C} and confirmed in \citet{2022A&A...664A..56S,2022A&A...666L..11S}. Here, we emphasise that these deep jets facilitate the formation of an advective adiabat at the same depths as \citet{2019A&A...632A.114S} propose to explain the inflated radius of HD209458b.  \\
In turn, strong latitudinal and vertical flows also develop, as can be seen in the meridional mass streamfunction (i.e. the meridional circulation profile - Equation 16 of \citealt{2019A&A...632A.114S}). In \autoref{fig:meridional_streamfunction} we plot the meridional circulation profile for the nominal model at near steady-state, with clockwise circulations shown in red and anticlockwise circulations shown in blue. Here we find that, at the equator, the strong stellar irradiation on the day-side leads to a general upwelling between $10^{-5}$ and $\sim1$ bar - driven by the combination of a clockwise circulation in the northern hemisphere and an anti-clockwise circulation in the south, both of which also drive material away from the substellar point/equator in the outer atmosphere. However, as we move deeper into the atmosphere, where the radiative time-scale is longer and hence advective effects can start to play a more significant role, we find that the sense of the meridional circulations has changed, likely due flows associated with the super-rotating jet taking over the vertical driving, leading to a strong downflow at the equator balanced by mass-conserving upflows at mid-latitudes (i.e. around $45^{\circ}$ - i.e. at the edge of the super-rotating jet). A similar circulation pattern was found by \citet{2021A&A...656A.128S} for Kepler-13Ab, a hot brown dwarf with a very hot (A-class) host star, and was shown to be sufficient to drive significant deep heating. \\

We next explore if this is also the case for our WASP-76b models. Specifically,
we start by investigating the vertical transport of enthalpy. We first recall briefly how this quantity
impacts the averaged energy transport in the atmosphere. Assuming the density is
near steady-state (a similar assumption to the anelastic approximation), the mass and energy conservation
equations are given by
\begin{eqnarray}\label{eq:steady_state}
  \bm{\nabla}\left(\rho \bm{u}\right) = 0, \cr
 \partial_t(\rho E)+ \bm{\nabla}\left((\rho e+\rho u^2/2+P+\rho \phi) \bm{u} +\bm{F_\mathrm{rad}}\right) = 0.
\end{eqnarray}
where $\rho$, $p$, $e$, and $E$ are the atmospheric density, pressure, internal,
and total
energy; $\bm{u}$ the velocity of the flow; $\phi$ the gravitational potential
and $\bm{F_\mathrm{rad}}$ the radiative flux (including the irradiation from the
host star). We will assume that the flow is
low Mach in the deep atmosphere and therefore neglect the contribution of the
kinetic energy. Furthermore we rewrite the energy flux as a function of the
enthalpy $\rho e +p = \rho c_p T$, with $c_p$ specific heat capacity at constant
pressure and $T$ the temperature. By averaging \autoref{eq:steady_state} in 2D over the full sphere ($\Omega$), we get 
\begin{eqnarray}
  \frac{1}{4\pi}\int_{4\pi} \rho u_z d\Omega = 0, \cr
  \partial_t\left(\frac{1}{4\pi}\int_{4\pi}\rho E d\Omega \right)+\partial_z\left(\frac{1}{4\pi}\int_{4\pi} \left(\rho c_pT+\rho \phi) u_z + F_\mathrm{rad}\right) d\Omega\right) = 0.
\end{eqnarray}
assuming there is no mass flux out of the domain of interest in a
plane-parallel approximation. The gravitational potential does not depend
on latitude/longitude, therefore, because of mass conservation, its contribution
to the energy flux is zero. Only the contribution of the enthalpy and the radiative flux remain:
\begin{eqnarray}
  \partial_t\left(\frac{1}{4\pi}\int_{4\pi}\rho E d\Omega \right) +\partial_z\left(\frac{1}{4\pi}\int_{4\pi} \left(\rho c_pT(z,\theta,\phi)u_z +
  F_\mathrm{rad} \right)d\Omega\right) = 0.
\end{eqnarray}
If the temperature is uniform, e.g. a 1D model, the contribution of the enthalpy is zero
similarly to the contribution of the gravitational potential. If not, e.g. a 3D
GCM, cold downflows and hot upflows will tend to cool the deep atmosphere whereas hot
downflows and cold upflows will tend to warm the deep atmosphere. This is how
the circulation can transport energy from the irradiated hot top layers to the
deep atmosphere, even in the absence of convective processes.\\

This split between upflows and downflows can be seen in \autoref{fig:enthalpy}, where we plot the longitudinal variation of the latitudinal-mean vertical enthalpy (top) and the horizontal-mean vertical enthalpy (bottom) for three models, two of which are near-steady-state (left - 1000 K initialisation - centre - 4000 K, i.e. nominal, initialisation) and one which was initialised with a hot deep adiabat that is still rapidly cooling at the time of the snapshot (right - 2500 K initialisation after 200 days). \\
Starting with the longitudinal variations of the latitudinal-mean vertical enthalpy (top - \autoref{fig:enthalpy}), it is clear that the direction of enthalpy transport varies significantly across the planetary surface. This was to be expected as tidally-locked thermal and wind dynamics, particularly in the outer atmosphere, are highly spatially inhomogeneous. However, this is further complicated by the effect that the temperature of the initial deep adiabat has on the overall dynamics - when a model is initialised with a deep atmosphere that is hotter than its final steady-state, excess energy must leave the deep atmosphere and, since radiative time-scales in the deep atmosphere are long, this typically occurs via changes in the wind structure and hence vertical enthalpy transport. An example of this effect can be found when comparing the models shown in \autoref{fig:enthalpy}: for the hot-start (2500 K) model near initialisation, \autoref{fig:Enthalpy_Profiles_2500K_slices}, we find that vertical enthalpy transport is primarily outwards, other than over a limited longitude and latitude range associated with a mass-conserving downflow. Almost the exact opposite scenario is found for a cool (1000 K) initialisation model (throughout its runtime), \autoref{fig:Enthalpy_Profiles_1000K_slices}, where we find that the enthalpy flow is directed downwards at most longitudes, albeit, once again, with a mass conserving counter flow. Finally the nominal model, \autoref{fig:Enthalpy_Profiles_nominal_slices}, represents a mix of the two regimes, with dynamics that can be linked to a combination of its very hot initialisation, leading to significant initial cooling, and long-run-time, leaving the model close to steady-state (although still warming in the deeper regions of the atmosphere due to the very-long dynamical times required to heat high-pressure regions of a hot Jupiter - see the isothermal model of \citet{2022A&A...658L...7K}). \\

This difference in regime is also reflected in the horizontal-mean vertical enthalpy profiles (bottom - \autoref{fig:enthalpy}): both the 1000 K and nominal models reveal a net downwards enthalpy flux, extending from the outer atmosphere all the way to the bottom of the simulation domain. Furthermore, this peak in the downwards flux is married with the radiative flux (both outwards and inwards) tending towards zero, as required in the potential temperature advection mechanism (\citealt{2017ApJ...841...30T}).   Note that the vertical extent of the enthalpy downflow is reduced in the nominal model when compared with the 1000 K model, which is due to the nominal model being closer to steady-state and hence heating being limited to the deepest regions of the simulation domain (see Figure 7 of \citealt{2019A&A...632A.114S} for an example of this top-down evolution - similar top down heating can be found in the 1400 K model as it warms back up from the initial cooling that occurred during model initialisation). {This effect (i.e. a switch from radiative to advective dynamics) can also be seen when comparing the vertical-advective and global-mean-radiative timescales: as we move deeper into the atmosphere, the dynamics switch from being radiatively dominated to advectively driven, at around the same pressure as the deep adiabat forms. However, it is important to note that this is a 1D view of an inherently 3D problem - between the tidally located nature of the planetary irradiation (i.e. the xied day-side and nigth-side), and the strong longitude and latitude dependence of the vertical winds, the exact pressure at which the atmosphere changes dynamical regimes is likely to be highly localised. Yet it is reassuring to confirm that, on a global scale, the regime transition occurs about where we would expect and as required for our mechanism to work.}
\\On the other hand, early outputs of the 2500 K model reveal, as expected, a strong enthalpy upflow throughout most of the deep atmosphere, although as the simulation evolves and the deep atmosphere finishes cooling, this slowly evolves towards the deep heating seen in the 1000 K and nominal models. Hints of this evolution towards deep heating can be seen around 1 bar where a weaker net downflow has started to develop. As shown in \autoref{tab:table1}, the global steady-state vertical enthalpy flux is generally independent of the initialisation temperature. That is to say that, given enough time, almost all the models here should settle onto the same steady-state profile, with the initialisation temperature only affecting the time taken to reach that profile. The only exception to this rule is the model in which we have modified the deep Rayleigh-drag. \\

As shown in \autoref{tab:table1}, the model with slower deep Rayleigh-drag exhibits a significantly stronger peak and importantly mean vertical enthalpy flux than the models with fast drag ($\tau_\mathrm{drag}=1$ day), even when models are compared at the same point in time ($\sim50,000$ days). This difference in vertical heating rate, and hence the temperature of the steady-state deep atmosphere, can be understood through using the vertical advective timescale ($\tau_\mathrm{adv}\sim\frac{H}{u_{r}}$): if we consider the scale height. $H$, to be on the order of the radius WASP-76b and the velocity to be the global mean vertical velocity ($u_r=734$ $\mathrm{m\,s^{-1}}$), we find that $\tau_\mathrm{adv}\sim2.06$ days {(see \autoref{fig:Time_Comp})}. I.e., for most of the models considered, the advective and drag time-scales in the deep atmosphere are of the same order of magnitude, leading to a noticeable reduction in the vertical wind speed, and hence vertical enthalpy flux in these models when comparing them with a no/low drag model in which the advective time-scale is significantly shorter than the drag time-scale (see \autoref{tab:table1}). A similar effect can be seen in the zonal-mean zonal-wind, with the Low Drag model exhibiting a jet that extends significantly deeper into the atmosphere than the nominal model it is based upon. We discuss the implications of this result on the expected level of advective radius inflation in \autoref{sec:Discussion}. \\

Finally we compare, in \autoref{fig:T_comp_1D_structure}, the near-steady-state temperature-pressure profile of the nominal model with both a 1D model of the outer and deep atmosphere calculated using ATMO (see \citealt{2015ApJ...804L..17T} for an overview of the ATMO model) and an internal-structure model (which extends down to over $10^{7}$ bar), based upon the work of \citet{2003A&A...402..701B,2004ApJ...603L..53C}, of a hot Jupiter with a mass of $0.9\mathrm{M_{J}}$. {This internal-structure model is rather unique, as it is very difficult to generate a model with such a large radius. In order to do so, a large amount of thermal energy (corresponding to a luminosity of $2\times10^{28}$ erg s$^{-1}$) must be deposited deep enough into the planetary interior to modify the internal adiabat (i.e. inflate the radius). As a consequence, the radius of the planet becomes essentially constant with time from early ages and the evolution is stalled (see Figure 4 of \citet{2004ApJ...603L..53C}).} \\
Note that the input physics and equation of state of these internal-structure models differs from that considered in expeRT/MITgcm (typically GCMs use simpler equations of state for computational efficiency reasons, and because they are focused upon relatively low-density dynamics). As such, the adiabatic index of our models and the internal-structure models also differ, complicating a direct comparison between the deep atmospheric temperature-pressure profiles in the two models. Instead, in order to divine which structure-model is the closet match to our steady-state GCM model, and hence calculate the level of radius inflation exhibited, we follow standard practice and perform the model comparison at a fixed pressure of 100 bar (i.e. at a reference-pressure which is sufficiently deep so that the atmosphere is optically thick and hence either convectively or advectively driven). \\
The result of this comparison is the selection of a internal-structure model with a radius of $R=1.98\mathrm{R_{J}}$ being chosen as the best `match' to our steady-state atmospheric model. This radius is broadly compatible with the observed radius of WASP-76b, $R=1.83\pm0.06\mathrm{R_{J}}$, suggesting that potential temperature advection alone is enough to explain the radius inflation of WASP-76b. A conclusion that is further reinforced by the partially evolved T-P profiles found in our alternate initialisation temperature models (see the solid lines in \autoref{fig:T_evo_comp}). Despite the shorter run time of the alternative start models, \autoref{fig:T_evo_comp} clearly shows that all of the models are converging towards the same, inflated, deep T-P profile found in the nominal model, albeit at different rates due to differences in the efficiency of deep cooling versus heating (see \citealt{2019A&A...632A.114S}), i.e. the slow heating of the 1000 K model in \autoref{fig:T_evo_comp}. This suggests that our conclusion of advection alone being sufficient to explain the radius inflation of WASP-76b is fairly robust.

\section{Discussion} \label{sec:Discussion}

In this work, we have performed additional analysis on extended and derivative versions of the WASP-76b models of \citet{2022A&A...666L..11S}, focusing our analysis on the vertical advection of potential temperature, and its ability to heat the deep atmosphere with respect to {\it 1D atmospheric models}, leading to radius inflation with respect to these 1D models (as introduced by \citealt{2017ApJ...841...30T} and explored, in a parametrised 3D model, by \citealt{2019A&A...632A.114S,2021A&A...656A.128S}). Importantly, thanks to the inclusion of a robust radiative transfer scheme (based upon petitRADTRANS) in expeRT/MITgcm, these models also allow us to complete the `wish' of \citet{2019A&A...632A.114S}: exploring the steady-state atmosphere of a hot Jupiter with a self-consistent radiative transfer scheme (in the outer atmosphere) so that a comparison between a atmospheric model and an internal-structure model can be made, thus quantifying, almost, the exact level of radius inflation that potential temperature advection alone can explain. \\

We started by exploring the zonal-mean zonal and meridional dynamics (\autoref{fig:nominal_zonal_means}), with the aim of confirming the presence of a strong super-rotating jet that drives an equatorial downflow between the irradiated outer atmosphere and the advective deep atmosphere. This analysis was performed for six models, five of which have different initial deep adiabat temperatures ranging from significantly hotter to cooler than the expected steady-state deep atmosphere (see the dashed lines in \autoref{fig:T_evo_comp}), and one which extends the nominal model of \citet{2022A&A...666L..11S}, but with slower deep Rayleigh-drag, and which we include in order to explore the robustness of our results. \\
For all five WASP-76b models with varying deep initialisation temperatures, we found that, once any deep atmospheric cooling had slowed/stopped, the strong super-rotating jet extends to $P>1$ bar and drives a meridional circulation profile that includes a zonal-mean downflow that connects the radiative outer atmosphere with the advective deep atmosphere. This implies that high potential temperature fluid parcels from the outer atmosphere can indeed be transported vertically downwards, potentially heating the deep atmosphere. \\

Next, we explored if this was indeed the case, investigating how the mean vertical enthalpy advection ($F_{\mathcal{H}r}\left(r,\theta,\phi\right) = \rho c_{p}T{U_r}\left(r,\theta,\phi\right)$) varies with both longitude and pressure (see \autoref{fig:enthalpy} and \autoref{tab:table1}). This analysis revealed a number of trends which line up with the dynamics of the atmosphere. For example, for models that are initialised with an overly hot deep adiabat, and hence exhibit significant initial, deep cooling, the primary direction of enthalpy transport is from the deep to the outer atmosphere where it can be radiated away. However as such a model evolves, and the deep atmosphere cools towards (and maybe overshoots - an effect seen in the hot initialisation models of \citealt{2021A&A...656A.128S}) steady-state, we find that all of our models exhibit a net downwards flow of enthalpy, with the strength and pressure range of the downwards transport decreasing as the deep atmosphere very slowly equilibrates (a process that can take many hundreds to thousands of Earth years for $P>100$ bar; \citealt{2019A&A...632A.114S,2021A&A...656A.128S}). Of course that is not to say that the vertical enthalpy transport lacks horizontal structure. As with the wind that drives it, differences in the vertical enthalpy transport are primarily linked with the differences in the day-side and night-side forcing, leading to a near global overturning circulation pattern that drives upwards vertical enthalpy transport on the day-side and downwards transport on the night-side, where divergent and wave driven circulations converge. We intend to explore the structure of the horizontal and vertical wind and enthalpy flux in more detail as part of a future study, including investigating how rotation impacts the dynamics (and hence may effect which hot Jupiters are inflated and which are not). \\
Overall we find that, regardless of the initial conditions (i.e. with enough time), all of our fast drag models exhibit comparable peak and mean vertical enthalpy transport into the deep atmosphere. Furthermore this vertical enthalpy transport is also comparable, if not slightly stronger than that found in a reanalysis of the HD209458b models of \citet{2019A&A...632A.114S}, reinforcing the idea that vertical potential temperature advection alone can explain the inflated radii of highly irradiated, gaseous, exoplanets. \\

We further confirm that this is the case via a comparison of our nominal WASP-76b models near-steady-state T-P profile (a T-P profile that all WASP-76b models appear to be converging towards - see \autoref{fig:T_evo_comp} - albeit at varying rates due to differences in the efficiency of cooling versus heating in the deep atmosphere) with an internal-structure model based upon the work of \citet{2003A&A...402..701B,2004ApJ...603L..53C}. As shown in \autoref{fig:T_comp_1D_structure}, the closest match to the nominal model is an internal-structure model with a mass of $0.9\mathrm{M_{J}}$ and an inflated radius of $1.98\mathrm{R_{J}}$, which is more than large enough to fully explain the observed radius of WASP-76b ($R=1.83\pm0.06\mathrm{R_{J}}$). Note however that this comparison was performed by only considering the temperature at 100 bar (a fairly standard pressure at which atmospheric and internal-structure model comparisons are performed), a necessary approximation given the rather different adiabatic indexes found in our models and the internal-structure models considered here. Briefly, this difference occurs due to differences in the physics and specifically the equation of state considered in the models, with expeRT/MITgcm using a relatively simplified EOS (for both computational efficiency reasons as well as the GCMs focus upon modelling relatively low-density regions of the atmosphere) in comparison to that used in \citet{2003A&A...402..701B,2004ApJ...603L..53C}. As such, an exact calculation of the level of radius inflation found in our model is beyond the current generation of GCMs, although work is in the pipeline to develop next-generation GCMs with updated dynamics and physics that will allow for even more robust comparisons with internal-structure models. However this does not mean that our calculation is without value, or that our results are far from the exact radius of our atmospheric model. For example, an internal-structure model with $\mathrm{R=R_{J}}$ is a very poor fit to our atmospheric model with deep temperatures at 100 bar that are a order of magnitude cooler than than found with expeRT/MITgcm, reinforcing our inference that this model exhibits significant, advectively driven, radius inflation. \\

However this is the not only effect that drives uncertainty in the exact level of radius inflation that advective heating can drive.  For example, \citet{2019ApJ...871...56M}, showed that the dynamics of small-Neptunes and super-Earths varied significantly between models which solved the primitive equations of meteorology the the full Navier-Stokes equations. Other model choices can also affect the strength of the deep heating, such as the strength of any grid-scale smoothing {(i.e. the inclusion of a Shapiro filter, which can affect the strength of the zonal jet and hence the vertical wind and advection - see \citealt{Koll_2018,2021MNRAS.504.5172S,10.1093/mnras/stac228})}, the atmospheric chemistry considered (e.g. equilibrium vs non-equilibrium chemistry){, or the sources of opacity included (for example the inclusion of SiO, Fe and FeII opacity may affect atmospheric heating and the depth to which radiation penetrates, changing the T-P profile slightly. See, for example \citealt{Lothringer_2020})}. Here we investigate one of these possible sources of uncertainty: the inclusion, and thus strength, of deep Rayleigh-drag. \\
This uncertainty can be seen by comparing the nominal model, with $\tau_\mathrm{drag}=1.0$ days, with a model in which the deep Rayleigh-drag has been significantly slowed, such that $\tau_\mathrm{drag}=1000$ days. i.e. a model in which the drag time-scale is significantly slower than the vertical advective timescale, which is of the order of 2 days for WASP-76b. Starting with the zonal-mean zonal-wind, our analysis indicates that the equatorial jet extends significantly deeper than in the nominal model. In turn, this drives stronger vertical mixing which results in a vertical enthalpy flux that is notably enhanced with respect to the nominal model. If we them compare the nominal model after 50,000 days with the low drag model after 40,000+10,000 days, we find that the deep T-P profile in the slow drag model is a little warmer, suggesting a slightly larger inflated radius. Comparing the vertical enthalpy flux at this time, confirms that the low drag model exhibits significantly enhanced deep heating. As such, and without a more complete understanding of how much, if any, Rayleigh-drag should be included in the deep atmosphere of hot Jupiter models, there will always remain an uncertainty on the exact level of radius inflation that vertical advection can drive. However, given that a) the Rayleigh-drag is confined to the highest pressure regions of the atmosphere (allowing for advective heat transport into the outer deep atmosphere, and then adiabatic mixing to carry heat deeper), and b) that the strength of the vertical advective transport is more than enough to explain the observed radius inflation, even in the nominal model with `strong' drag, we are confident in our conclusion that the vertical advection of potential temperature alone is enough to explain the radius inflation of many hot Jupiters (and hot brown dwarfs), including WASP-76b. \\

\section{Concluding Remarks}

Overall, our analysis of the vertical mixing and vertical transport of potential temperature in an extended sample of the the WASP-76b models of \citet{2022A&A...666L..11S} has revealed that, contrary to their conclusions, the vertical advection of potential temperature alone is more than enough to explain the radius inflation of WASP-76b. \\
This difference in conclusion arises for a number of reasons. \\
The first is simply that the nominal model of \citet{2022A&A...666L..11S} was not run for long enough, and that their approach to avoid the computational expense of evolving a radiative GCM to steady-state in the deep atmosphere (i.e. the steroids model) made a number of assumptions about the deep dynamics which limit the applicability of such a extrapolative approach. Specifically, when extrapolating the evolution of their nominal models deep P-T profile, they focused on the evolution of the temperature at 650 bar, which, for the time frame they considered, revealed near exponential cooling. However, as shown in the isothermal-start model of \citet{2019A&A...632A.114S}, advective heating of the deep atmosphere starts in the lower pressure regions of the deep atmosphere (i.e. at the bottom of the radiatively dominated outer atmosphere) and slowly pushes deeper with time, with the time to heat the atmosphere only increasing as the heating moves deeper and the local density increases. Evidence for this top down heating in the steroids models can be seen in Figure 2 of \citet{2022A&A...666L..11S}, with slow heating occurring between $\sim5$ and $\sim100$ bar, leaving the region around $650$ bar to appear steady and hence evolved. Here, by evolving the nominal model for an additional 69000 days of simulation time, we are approaching a true steady-state that is significantly hotter than the steroids model. Furthermore, when compared with an internal-structure model from \citet{2003A&A...402..701B,2004ApJ...603L..53C}, this steady-state corresponds to a radius of $1.98\mathrm{R_{J}}$, more than large enough to explain the observed, inflated, radius of WASP-76b ($R=1.83\pm0.06\mathrm{R_{J}}$). \\
The second reason for our difference in conclusion can be linked to the wide use of intrinsic/internal temperatures in the exoplanetary communities. Briefly, radius inflation is simply the difference between the observed radii of a hot Jupiter and a standard `radiative-convective' 1D model its outer atmosphere. This difference is believed to occur because 1D atmospheric models lack some fundamental physics that drive deep heating, with suggestion ranging from ohmic dissipation to vertical heat transport, and is `fixed' (or accounted for) in 1D models by including an artificial, intrinsic/internal temperature meant to represent the heating of the deep atmosphere. Commonly this is linked with excess energy loss from the interior (hence the name internal temperature), however \citet{2017ApJ...841...30T} and \citet{2019A&A...632A.114S} proposed that this deep heating instead occurs due to vertical heat transport, with no need for any energy transport from the interior to the outer/deep atmosphere (i.e. zero net deep flux). In essence, this intrinsic temperature acts as a `fudge' factor designed to allow for direct comparisons between observations (such as transmission spectra) and 1D models, and relying upon it outside of those scenarios can lead to either over or under (as was the case in \citealt{2022A&A...666L..11S}) estimation of the level of radius inflation. Instead, as done here, comparisons must be done with internal-structure models, even when the accuracy of those comparisons is limited by the different equations of state used (i.e. by the simplified EOS used in GCMs - although work is in progress to change this).\\

Of course, many questions remain about the exact level of radius inflation that vertical advection can drive, and if it can fully explain the differences seen in radius inflation for the broader hot Jupiter community, including those unusual objects that are very highly irradiated and yet show little to no sign of inflation (for example WASP-43b or WASP-18b). expeRT/MITgcm now makes a radiatively robust study of these objects possible for the first time, and we look forward to the results of future work with this, and other next-generation, models. \\
{However there is now no doubt that potential temperature advection provides a robust explanation for some if not all of the observed radius inflation of hot Jupiters and hot brown dwarfs, and as such changes to how future GCM studies are performed are recommended. Previously, it has been recommended that future GCM studies of hot Jupiters be initialised with a adiabat at the bottom of their simulation domain and then be allowed to evolve to steady-state \citep{2019A&A...632A.114S}. However this remains computationally expensive and can lead to mistakes when models are not allowed to evolve sufficiently. As such, given how well potential temperature advection alone can explain the inflated radii of hot Jupiters, we suggest that future studies should initialise their deep atmosphere with an adiabat based upon the best fitting internal-structure model that corresponds to the inflated radii, albeit modified to match the adiabatic index of the GCM.  }

\section*{Acknowledgements}
We would like to thank the anonymous referee for their comments and corrections.
F. Sainsbury-Martinez would like to thank UK Research and Innovation for support under grant number MR/T040726/1
I.B receives support from  the ERC grant No. 787361-COBOM and the consolidated STFC grant ST/R000395/1.
A.D.S., L.D., U.G.J
and C.H. acknowledge funding from the European Union H2020-MSCA-ITN-2019 under Grant no. 860470
(CHAMELEON). L.C. acknowledges the Royal Society University Fellowship
URF R1 211718 hosted by the University of St Andrews. U.G.J acknowledges
funding from the Novo Nordisk Foundation Interdisciplinary Synergy Program
grant no. NNF19OC0057374.

\section*{Data Availability}
Data is available upon request. 



\bibliographystyle{mnras}
\bibliography{papers} 








\bsp	
\label{lastpage}
\end{document}